\documentstyle[12pt]{article}

\begin{document}

\def\onehalf{{\textstyle \frac12}}
\def\tsty#1#2{{\textstyle\frac{#1}{#2}}}
\def\trans{^{\scriptscriptstyle\top}}
\def\ie{{\it i.e.}}
\def\pmbf#1{\rlap{$#1$}{}\hskip-0.5pt{#1}}
\def\abs#1{{\scriptstyle|}#1{\scriptstyle|}}
\def\aabs#1{{\scriptscriptstyle|}#1{\scriptscriptstyle|}}
\def\of#1{{{\scriptstyle(}#1{\scriptstyle)}}}
\def\oof#1{{{\scriptscriptstyle(}#1{\scriptscriptstyle)}}}
\def\brof#1{{{\scriptstyle[}#1{\scriptstyle]}}}
\def\broof#1{{{\scriptscriptstyle[}#1{\scriptscriptstyle]}}}
\def\totder#1#2{\frac{d #1}{d #2}}
\def\parder#1#2{\frac{\partial #1}{\partial #2}}
\def\Poi#1#2{\{#1,#2\}}
\def\ssr#1{{\scriptscriptstyle\rm #1}}
\def\ssty#1{{\scriptscriptstyle #1}}
\def\vecdos#1#2{\left({\matrix{#1\cr #2\cr}}\right)}
\def\matdos#1#2#3#4{\left({\matrix{#1&#2\cr #3&#4\cr}}\right)}
\def\vectres#1#2#3{\left(\matrix{#1 \cr #2 \cr #3} \right)}
\def\mattres#1#2#3#4#5#6#7#8#9{\left(\matrix{#1&#2&#3\cr #4&#5&#6\cr
#7&#8&#9\cr} \right)}
\def\figbox{\framebox{\hbox to5cm{\vbox to1cm{\vfil}\hfil}}}
\def\Figura#1{\bigskip\bigskip\vbox{\hrule\bigskip{\small\noindent{\bf
Figure }{#1}}\bigskip\hrule}\bigskip\bigskip}
\def\jour#1#2#3#4{{\sl #1{}} {\bf #2}, #3\ (#4)}
\def\ket#1{\,\vert{#1}\rangle}
\def\bra#1{\langle{#1}\vert}
\def\braket#1#2{\langle{#1}\vert{#2}\rangle}
\def\rket#1{\,\vert{#1})}
\def\rbra#1{({#1}\vert}
\def\rbraket#1#2{({#1}\vert{#2})}
\def\ketdos#1#2{\left\vert{#1\atop #2}\right\rangle}
\def\brados#1#2{\left\langle{#1\atop #2}\right\vert}
\def\su{\hbox{su(2)}}
\def\so{\hbox{so(2)}}
\def\suq{\hbox{su$_q$(2)}}

\newcommand{\be}{\begin{equation}}
\newcommand{\ee}{\end{equation}}
\newcommand{\bea}{\begin{eqnarray}}
\newcommand{\eea}{\end{eqnarray}}


\noindent {\Large \bf Finite {\it q\/}-Oscillator}
\\[1cm]

{\bf Natig M.\ Atakishiyev,\footnote{Instituto de Matem\'aticas,
UNAM, Apartado Postal 273-3, 62210 Cuernavaca, Morelos, M\'exico}
Anatoliy U.\ Klimyk,$^{1,}$\footnote{Bogolyubov Institute for
Theoretical Physics, Kiev 03143, Ukraine}

and Kurt Bernardo Wolf} \\[8pt]

{Centro de Ciencias F\'{\i}sicas, UNAM, Apartado Postal 48-3,
62251

 Cuernavaca, Morelos, M\'exico}

\bigskip

\bigskip

\begin{abstract}

The finite $q$-oscillator is a model that obeys the dynamics of
the harmonic oscillator, with the operators of position, momentum
and Hamiltonian being functions of elements of the $q$-algebra
\suq.  The spectrum of position in this discrete system, in a
fixed representation $j$, consists of $2j{+}1$ `sensor'-points
$x_s=\onehalf[2s]_q$, $s\in\{{-j},{-j}{+1},\ldots,j\}$, and
similarly for the momentum observable. The spectrum of energies is
finite and equally spaced, so the system supports coherent states.
The wave functions involve dual $q$-Kravchuk polynomials, which
are solutions to a finite-difference Schr\"odinger equation. Time
evolution (times a phase) defines the fractional
Fou\-rier-$q$-Kravchuk transform. In the classical limit $q\to1$
we recover the finite oscillator Lie algebra, the $N=2j\to\infty$
limit returns the Macfarlane--Biedenharn $q$-oscillator and both
limits contract the generators to the standard quantum-mechanical
harmonic oscillator.

\end{abstract}

\bigskip


\vfil

\eject

\noindent
{\bf 1. Introduction}
\bigskip

    Discrete models which are counterparts to well-known continuous
systems, and in particular those which contract to the standard
harmonic oscillator, are of fundamental interest in theoretical
physics \cite{AtSus}--\cite{ArikAW}. Moreover, {\it finite\/}
discrete models are also of interest for the parallel processing
of signals through nano-optical devices, where the input may
involve lasing carbon tubules, the output being registered by a
finite sensor array, and the device consisting of a shallow planar
waveguide---an oscillator which can carry only a finite number of
states \cite{AW97}. The salient purpose of such a device is to
perform a finite analogue of the fractional Fourier transform
\cite{Ozak}.

    In previous works on the one-dimensional finite oscillator
\cite{AtWo,AW97,AVW99}, {\it oscillator\/} systems were characterized
in the familiar context of Hilbert spaces and Lie algebraic theory;
in \cite{ArikAW} these requirements were formalized into the three
following postulates:

\medskip

\noindent{\bf 1.}\quad There exists an essentially self-adjoint
{\it position} operator, indicated $Q$, whose spectrum $\Sigma(Q)$
is the set of positions of the system.

\medskip

\noindent{\bf 2.}\quad There exists a self-adjoint and compact
{\it Hamiltonian} operator, $H$, which generates time evolution
through the Newton-Lie, or equivalent Hamil\-ton-Lie equations:
\be
    [H,[H, Q]]= Q \quad\Longleftrightarrow\quad
    \left\{ \begin{array}{rcl}
        [H, Q] &=:& -{\rm i} P,\\ {}
         [H, P] &=& \phantom{-}{\rm i} Q, \end{array} \right.
                \label{eq:Newton-Hamilton}
\ee where $[\,{\cdot}\,,\,{\cdot}\,]$ is the commutator. The first
Hamilton equation in (1) defines the {\it momentum} operator $P$,
while the second one contains the harmonic oscillator dynamics.
The  set of momentum values of the system is the spectrum
$\Sigma(P)$ of $P$.

\medskip

\noindent{\bf 3.}\quad The three operators, $Q$, $P$ and $H$,
close into an {\it associative algebra}, \ie, satisfy the Jacobi
identity, \be
    [P,[H,Q]]+[Q,[P,H]]+[H,[Q,P]]=0.  \label{eq:Jacobi-identity}
\ee

    The second and third postulates determine that $[Q,P]$ must commute
with $H$, which implies that it can only be of the form $[Q,P]={\rm i}\,F(H)$,
where $F$ is some function of $H$ (including constants) and the $\rm i$
is placed to make $F(H)$ self-adjoint, but do not otherwise specify this
basic commutator further. For a constant $F(H)=\hbar{\hat1}$, one recovers
the standard oscillator algebra $H_4=\hbox{span}\,\{H,Q,P,\hat1\}$, which
contains the basic Heisenberg-Weyl subalgebra $W_1=\hbox{span}\,\{Q,P,\hat1\}$
of quantum mechanics. In our first works \cite{AW97,AVW99} we examined
the cases which, in the unitary irreducible representations of spin
$j=\onehalf N$ ($N\in\{0,1,\ldots\}$ fixed), correspond to the linear
function $F(H)=H-(j+\onehalf)\hat1=:J_3$, and so the operators close into
the  Lie algebra ${\rm so}(3)={\rm su}(2) = \hbox{span}\,\{Q,P,J_3\}$.
The purpose of the present paper is to study the case when, for $q:=e^{-\kappa}$,
the basic commutator is
\bea
    {}[Q,P]&=&{\rm i}\,F_q(H),\qquad H=J_3+(j+\onehalf)\hat1,
                                \label{eq:f-of-suq0}\\
        F_q(H)&=& e^{-2\kappa J_3}
            \frac{\cosh\onehalf\kappa}{2\sinh\onehalf\kappa}
            -e^{-\kappa J_3}
            \frac{\cosh(j+\onehalf)\kappa}{2\sinh\onehalf\kappa}
                        \label{eq:f-of-suq}\\
            &=&\onehalf e^{-\kappa J_3}
                \Big(e^{-\kappa J_3}\cosh\onehalf\kappa
                    - T_{2j+1}(\cosh\onehalf\kappa)\Big)
                        \Big/ \sinh\onehalf\kappa,
                        \label{eq:f-of-suq-wT}
\eea
where $T_n$ is the Chebyshev polynomial of the first kind, and
$q\in(0,1]$ or $\kappa\in[0,\infty)$. In particular, $F_1(H)=J_3$
returns the previous ${\rm su}(2)$ case \cite{AW97}.

    An important ingredient for the postulates of harmonic oscillator
dynamics is an unambiguous correspondence between the physical
observables of position, momentum and energy, with the elements of
the associative algebra. In Section 2 we recall the main relevant
results on the algebra \suq\ and its standard representation
basis. The \suq\ nonstandard basis, investigated in
\cite{AW00,AK02}, is introduced in Section 3 to exhibit our
proposed correspondence explicitly in terms of the generators of
\suq. With our postulated choice, the position and energy spectra
in the $(2j{+}1)$-dimensional representation $j=\onehalf N$ of
\suq\ will be \bea
    \Sigma(Q) &=& x_s = \onehalf[2s]_q
            =\frac{\sinh s\kappa}{2\sinh\onehalf\kappa},
                  \quad s\in\{-j,-j{+}1,\ldots,j\}=:s|_{-j}^j,
                            \label{eq:position-energy-spectra1}\\
    \Sigma(H) &=& E_n = n+\onehalf,
                   \qquad n\in\{0,1,\ldots,2j\}=:n|_0^{2j}.
                            \label{eq:position-energy-spectra2}
 \eea
We recall the definition of the $q$-number for
$q=e^{-\kappa}$: \be
    \brof{r}_q=\brof{r}_{q^{-1}}=-\brof{{-r}}_q
        :=\frac{q^{\frac12r}-q^{-\frac12r}}{q^{\frac12}-q^{-\frac12}}
         = \frac{\sinh\onehalf r\kappa}{\sinh\onehalf\kappa}.
                                        \label{eq:def-q-box}
\ee
Note that the $q$-number of an integer $r$ is
$U_{r-1}(\cosh\onehalf\kappa)$, the Chebyshev polynomial
of the second kind. The spectrum of momentum is the same as that
of position, $\Sigma(P)=\Sigma(Q)$. The classical limit is
$\lim_{q\to1}\brof{s}_q=s$, when the $q$-algebra \suq\ becomes the
Lie algebra \su; then, the set of positions become equally spaced
and we are back at the previously known finite oscillator \cite{AW97}.
But for all other values of the deformation parameter $q$, the `sensor
points' of the system are concentrated towards the center of the
interval, while the endpoints are spread farther apart. Yet the
energy spectrum remains an equally-spaced set, and therefore the
system follows harmonic motion.

    The finite $q$-oscillator wave functions are the overlaps between
the position and energy eigenbases. They are written out in Section 4
in terms of the dual $q$-Kravchuk polynomials, and are orthonormal
and complete over the sensor points of the system. The momentum
representation of these wave functions is addressed in Section 5 with
the Fourier-$q$-Kravchuk transform, and in Section 6 this transform
is fractionalized. The evolution in time of a finite $q$-oscillator
(or equivalently, the parallel processing of a finite signal along
the axis of a shallow planar waveguide), is the 2-fold cover of the
fractional Fourier-$q$-Kravchuk transform matrix; the metaplectic
sign appears thus for half-integer values of $j$, which corresponds
to a finite systems of an even number of points. In Section 7 we
introduce the concept of an {\it equivalent potential\/} for discrete
systems which is based, as in the continuous case, on the existence
of a ground state with no zeros.  Finally, in Section 8 we verify
that the contraction limits $q\to1$ and $N=2j\to\infty$ of the
algebra \suq\ reproduce the known results for the finite oscillator
and the continuous $q$-oscillator. The corresponding limits for the
wave functions however, present further challenge.

\bigskip

\noindent
{\bf 2. The algebra \suq\ and its standard basis}
\bigskip

    The quantum algebra \suq\ is the associative algebra
generated by three elements, usually denoted as $J_+,J_-,J_3$,
subject to the commutation relations
\be
    [J_3,J_{\pm}] = \pm J_{\pm},\qquad [J_+, J_-] = [2 J_3 ]_q.
                            \label{eq:commrel-Jpm3}
\ee
Equivalently, writing $J_\pm=J_1\pm {\rm i}J_2$, we characterize the
algebra \suq\ by
 \be
    [J_2,J_3]={\rm i}\,J_1,\quad [J_3,J_1]={\rm i}\,J_2,\quad
{\textstyle    [J_1,J_2]=\frac {\rm i}2 [2J_3]_q.     }
\label{eq:suq-comm-rel}
 \ee
The first two commutators in (\ref{eq:suq-comm-rel}) have the structure
of the oscillator Hamilton equations (\ref{eq:Newton-Hamilton}), while
the third one involves the $q$-number (\ref{eq:def-q-box}), which
distinguishes $q$-algebras from Lie algebras, the latter corresponding
to the case $q=1$. The following element in the covering algebra of
\suq\ commutes with all others,
\be
    \begin{array}{rcl}
    C_q&:=& J_1^2+J_2^2
        + [J_3{-}\onehalf]^2_q+\onehalf[2J_3]_q-\tsty14 \\[3pt]
        &=&J_+J_-+[J_3-\onehalf]_q^2-\tsty14,  \end{array}
                    \label{eq:def-Casimir-op}
\ee
and is called its {\it Casimir\/} operator.

    It is convenient to have a realization of the \suq\ generators
in terms of first-degree differential operators, acting on spaces
${\cal H}_j$ of functions of a formal variable $x$, and depending
on the numerical irreducible representation label $j$. This is
\bea
    J_+&:=&J_1+ {\rm i}\,J_2\quad{\leftrightarrow}\quad
            x\Big[2j-x\totder{}{x}\Big]_q
            =x\,[j-J_3]_q,\label{eq:Jplus}\\
    J_-&:=&J_1-{\rm i}\,J_2\quad{\leftrightarrow}\quad
            \frac1x\Big[x\totder{}{x}\Big]_q
            =\frac1x[j+J_3]_q,\label{eq:Jminus}\\
    J_3&\leftrightarrow& x\totder{}x - j,\qquad
                    j\in\{0,\onehalf,1,\ldots\}\hbox{ fixed}.
                                \label{eq:Jthree}
\eea

    The set of power monomials $x^{j+m}|_{m=-j}^j$ are eigenfunctions
of $J_3$ and provide the {\it standard\/} basis for the irreducible
space ${\cal H}_j$, of finite dimension $2j+1$. The functions of the
basis were chosen in \cite{AW00,AK02} with the following constants:
\be
    f_m^j\of{x} := q^{\frac14(m^2 - j^2)} \,
            \left [ {2j \atop j{+}m}\right ]_q^{1/2} \, x^{j+m},
                    \label{eq:def-standard-basis}
\ee
where the $q$-binomial coefficient $\left[{m\atop n}\right
]_q$ is defined (using the standard notation of $q$-analysis
\cite{GasRah}) for $m\ge n$ nonnegative integers by
\bea
    {}\left[{m\atop n}\right]_q
    &:=& \frac{(q;q)_m}{(q;q)_n(q;q)_{m-n}}
        =(-1)^nq^{mn-\frac12 n(n-1)}
            \frac{(q^{-m};q)_n}{(q;q)_n} , \label{eq:q-bin-coeff}\\
     (z;q)_n &:=& \prod _{k=0}^{n-1}(1-zq^k),\quad n=1,2,3,...,
            \qquad (z;q)_0 = 1.
                    \label{eq:q-bin-coeff2}
\eea
For any two complex vectors ${\bf a},{\bf b}\in{\cal H}_j$,
\be
     {\bf a}=\sum_{m=-j}^j\alpha_m f_m^j,\qquad
      {\bf b}=\sum_{m=-j}^j\beta_m f_m^j,
                \label{eq:arbitr-vect-AB}
\ee
there is a natural sesquilinear inner product
 \be
    ({\bf a},{\bf b})_{{\cal H}_j}:=\sum_{m=-j}^j\alpha_m^*\,\beta_m,
                    \label{eq:inner-product}
\ee
with respect to which the standard basis is orthonormal. The action
of the \suq\ generators and Casimir operator on the standard basis
is well known:
\bea
    J_3 f_m^j= mf_m^j,
    & & J_\pm f_m^j=\sqrt{[j\pm m+1]_q\,[j\mp m]_q}\,\,f_{m\pm1}^j,
                         \label{eq:J3pm-raise-lower}\\
    C_q f_m^j= c_q\,f_m^j,
    & & c_q :=[j{+}\onehalf]_q^2-\tsty14. \label{eq:c-eigenvalues}
\eea
These equations are of course independent of the realization of the
basis vectors $f_m^j$ by the power monomials $f_m^j\of{x}$ in $x$.

    The spectrum of the diagonal generator $J_3$ [see (\ref{eq:Jthree})
and (\ref{eq:J3pm-raise-lower})] is linear and bounded, as that of a
finite version of the quantum harmonic oscillator. Indeed, this is our
choice for the finite $q$-oscillator Hamiltonian, displaced so that the
ground state has energy $\onehalf$, namely
\be
    H= J_3+j+\onehalf, \qquad
                H\,f_m^j=(n+\onehalf)\,f_m^j,
                        \quad n:=j+m,    \label{eq:def-H-Jthree}
\ee
where $n|_0^{2j}$ is the {\it mode number\/} that counts the number
of energy quanta. At this point we are presented with what would appear
as a `natural' assignment for the position and momentum operators,
$Q\leftrightarrow J_1$ and $P\leftrightarrow -J_2$, because it would
be the simplest generalization of the previously studied $q=1$ case
\cite{AW97,AVW99}. This choice would bring the first two commutators
in (\ref{eq:suq-comm-rel}) to reproduce correctly the two Hamilton
equations in (\ref{eq:Newton-Hamilton}), while the third commutator
$[Q,P]$ would have the form (\ref{eq:f-of-suq0}) with $F_q(H)=
\onehalf[2J_3]_q = {\sinh\kappa(H-j-\onehalf)}/{2\sinh\onehalf\kappa}$.
In this `na\"{\i}ve' model however, the spectra of $Q$ and $P$ are not
algebraic; they must be computed numerically as roots of a polynomial
equation of degree $2j+1$.

\bigskip

\noindent
{\bf 3. The nonstandard basis}
\bigskip

    While we do not discard the model suggested at the end of the
previous Section, we find more attractive to propose a
correspondence between the physical observables of position and
momentum, $Q,P$, and the {\it nonstandard\/} (also called {\it
twisted\/}) operators (see \cite{Nom-1}--\cite{BCh99},
\cite{AW00,AK02}), which have the virtue of possessing an
algebraic spectrum $x_s:=\onehalf[2s]_q$, $s|_{-j}^j$. The
position (and hence momentum) observables will be thus identified
with the following operators: \bea
    Q&=& \widetilde J_1
        :=  q^{\frac14J_3}\,J_1\,q^{\frac14J_3},
                        \label{eq:Q-def-twiddle}\\
    -P&=& \widetilde J_2
        :=  q^{\frac14J_3}\,J_2\,q^{\frac14J_3},
                        \label{eq:P-def-twiddle}
\eea
while the Hamiltonian $H$ is associated to $J_3$ by
(\ref{eq:def-H-Jthree}) as before.

    We note that while the $q$-number (\ref{eq:def-q-box}) displays
symmetry under $q$-inversions, $q\leftrightarrow q^{-1}$,
$\brof{r}_q=\brof{r}_{q^{-1}}$, the identification of tilded operators
in (\ref{eq:Q-def-twiddle})--(\ref{eq:P-def-twiddle}) preserves this
symmetry with the concomitant reflection $J_3\leftrightarrow -J_3$.
This means that the ground state of a $q<1$ oscillator is the top
state of its $q^{-1}>1$ partner.

    The commutation relations among the nonstandard operators and
$J_3$ are
\bea
    && [J_3,Q]=-{\rm i}\,P,\qquad
       [J_3,P]={\rm i}\,Q,\label{eq:tJpm-comm-rels}\\
  {} [Q,P]
    &=&{\textstyle {\frac {\rm i}2}\, q^{\frac12J_3}(q^{-\frac12}J_+J_-
            -q^{\frac12}J_-J_+)q^{\frac12J_3}=:{\rm i}\,F_q(C_q,J_3)    }
                    \label{eq:def-J3twiddle}\\
    &=&{\rm i}\Big(e^{-\kappa J_3}[(C_q+\tsty14)\sinh\onehalf\kappa
            +\onehalf\hbox{csch}\,\onehalf\kappa ]
            -\onehalf \,e^{-2\kappa J_3}\,\coth\onehalf\kappa\Big),
                                                    \nonumber
\eea
where $q=e^{-\kappa}$ as before. The operator $F_q(C_q,J_3)$ defined
in (\ref{eq:def-J3twiddle}) commutes with $J_3$ and is also diagonal
in the standard basis; in the irreducible representation $j$,
\be
    F_q\,f_m^j  = \frac{e^{-2m\kappa}\cosh\onehalf\kappa
        -e^{-m\kappa}\cosh(j{+}\onehalf)\kappa}{2\sinh\onehalf\kappa}
                    \,f_m^j,    \label{eq:widetilde-J-diagonal}
\ee
but its spectrum is {\it not\/} a good candidate for an oscillator
Hamiltonian, because it is not equally spaced [unlike
(\ref{eq:position-energy-spectra2})], and so the motion would not
be harmonic, but dispersive. In terms of the position and momentum
generators (\ref{eq:Q-def-twiddle})--(\ref{eq:P-def-twiddle}), the
Casimir operator (\ref{eq:def-Casimir-op}) acquires the form
\bea
    C_q
   &=&\hbox{sech}\,\onehalf\kappa\, (Q^2+P^2)\,e^{\kappa J_3}
                        + D_q(J_3), \label{eq:def-Casimir-opQPJ}\\
    D_q(J_3)&:=&\hbox{sech}\,\onehalf\kappa\, \Big([J_3{-}\onehalf]_q^2
                -\onehalf e^{-\kappa J_3}\coth\onehalf\kappa
                +\onehalf\hbox{csch}\,\onehalf\kappa\Big)
                     -{\textstyle\frac14}.
                                    \label{eq:def-Dq(J)}
\eea

    We recall a previous phase-space picture for the finite oscillator
of $2j{+}1$ points, considered in  \cite{AChW98}, as the (classical)
sphere $Q^2+P^2+J_3^2=j(j{+}1)$, having circular sections of square
radius $Q^2+P^2=(j{+}\onehalf)^2-(J_3{-}\onehalf)^2-J_3$. For \suq,
the corresponding surface now has the section
\bea
        Q^2+P^2&=&\Big([j{+}\onehalf]_q^2\cosh\onehalf\kappa
                -[J_3{-}\onehalf]_q^2 \nonumber \\
            & &\qquad{}+\onehalf e^{-\kappa J_3}\coth\onehalf\kappa
                -\onehalf\,\hbox{csch}\,\onehalf\kappa\Big)\,
                        e^{-\kappa J_3} .
                            \label{eq:pictureof-phasespace}
 \eea
Phase space for the finite $q$-oscillator is suggested thus as
$q$-dependent pear-shaped sphero\"{\i}ds, tip-up for $q<1$ and
tip-down for $q>1$ (recall the $q\leftrightarrow q^{-1}$ symmetry
with the inversion of $J_3$). The $q$-harmonic oscillator
evolution (\ie, a phase times the so-defined fractional
$q$-Fourier-Kravchuk transform) will rotate this space around the
$J_3$ vertical symmetry axis of the sphero\"{\i}d.

     In this finite $q$-oscillator model we interpret the eigenvalues
$x_s$ of $Q:=\widetilde J_1$ as the discrete values of the position
observable. The eigenfunctions $g^j_s\of{x}$ and eigenvalues of this
nonstandard operator were found in \cite{AW00}, and they are of the form
\bea
    Q g^j_s\of{x}&=& x_s\,g^j_s\of{x},\quad
            x_s=\onehalf [2s]_q
                =\frac{\sinh s\kappa}{2\sinh\frac12\kappa}
                =-x_{-s},\quad s|_{-j}^j,
                        \label{eq:eigenv-eq-J1}\\
        g^j_s\of{x} &=& \gamma^j_s\,
            (q^{\frac14(1-2j)} x;q)_{j-s}\,
            (-q^{\frac14(1-2j)} x;q)_{j+s}=g^j_{-s}\of{{-x}},
                        \label{eq:eigenf-eq-g}\\
        \gamma^j_s&:=& q^{\frac12(j+s)}
                \sqrt{\bigg[{2j\atop j+s}\bigg]_{q^2}
           \frac{1+q^{-2s}}{2(-q;q)_{2j}}}.\label{eq:c-in-eq}
\eea
They are normalized with respect to the inner product
(\ref{eq:inner-product}), and are orthogonal because they correspond
to distinct eigenvalues $x_s$. This basis of $2j+1$ functions
$g^j_s\of{x}$, $s|_{-j}^j$ we call the {\it position\/} basis. A
signal consisting of $2j+1$ values $\Phi_s$, sensed at the positions
$x_s$ [given in (\ref{eq:position-energy-spectra1})], is
\be
        \Phi=\sum_{s=-j}^j \Phi_s\,g^j_s \in{\cal H}_j,
                        \label{eq:def-gen-abs-vector}
\ee
and can be realized either as a function of $x$, or as a
$(2j+1)$-dimensional column vector with components numbered by
$s|_{-j}^j$.

\bigskip

\noindent
{\bf 4. Finite $q$-oscillator mode wave functions}
\bigskip

    We have now two bases for ${\cal H}_j$: the standard basis
$\{f_m^j\}_{m=-j}^j$ of {\it mode\/} $n=j+m$ (and energy
$E_n=n+\onehalf$), and the nonstandard basis $\{g_s^j\}_{s=-j}^j$
of {\it position\/} $x_s=\onehalf [2s]_q$. In the realization of
\suq\ generators given in (\ref{eq:Jplus})--(\ref{eq:Jthree}), the
mode basis is realized by the power functions in
(\ref{eq:def-standard-basis}), and the position basis by
(\ref{eq:eigenf-eq-g}). We can use this realization to find the
unitary transformation between these two orthonormal bases, and
thus define the finite $q$-oscillator wave functions by the
overlap
\be
    \Phi^{{(2j|q)}}_n(x_s)
            :=(g^j_s,f^j_m)_{{\cal H}_j}\quad \left\{
        \begin{array}{ll}\hbox{of mode $n=j+m$},& n|_0^{2j},\\
        \hbox{on points $x_s=\onehalf[2s]_q$},& s|_{-j}^j.
                \end{array}\right. \label{eq:def-of-Phi}
\ee
By construction, this set of functions is orthonormal and complete
under the ${\cal H}_j$ inner product (\ref{eq:inner-product}).

    The overlap (\ref{eq:def-of-Phi}) is obtained by expanding the
function $g^j_s\of{x}$ of (\ref{eq:eigenf-eq-g}) into a power
series in $x$, which is then
\be
    g^j_s\of{x}=\sum_{m=-j}^j \Phi^{{(2j|q)}
                    }_{j+m}(x_s)^*\,f^j_m\of{x},
                \quad
    f^j_m\of{x}=\sum_{s=-j}^j \Phi^{{(2j|q)}
                    }_{j+m}(x_s)\,g^j_m\of{x}.
                \label{eq:expansion-fggf}
 \ee
The expansion of $g_s^j(x)$ in $x$ is \cite{AW00} \be
    g^j_s\of{x}= \gamma^j_s \sum _{m=-j}^j
            q^{\frac14(j+m)(j+m-1)}\,
                {\bigg[{2j\atop j{+}m}\bigg]_q}^{1/2}\,
                K_{j+m}(\lambda\of{j{-}s}; -1 ,2j\,\vert\,q)
                        \,f^j_{m}\of{x},  \label{eq:g-exp-f}
\ee
expressed in terms of the {\it dual $q$-Kravchuk polynomials},
\be
    K_n (q^{-\xi}{+}cq^{\xi-2j}; c, 2j | q)
    := {}_3\phi_2 \left({q^{-n},\,q^{-\xi},\, c q^{\xi-2j}\atop
             q^{-2j},\  0}\bigg |\, q\,; \,q \right),
                            \label{eq:def-dual-Kravchuk}
\ee
where ${}_3 \phi _2$ is the basic hypergeometric function defined in
\cite{GasRah}, and the coefficients $\gamma^j_s$ are given
in (\ref{eq:c-in-eq}).

    In the particular case of our concern, the argument of the dual
$q$-Kravchuk polynomial is $\lambda(\xi)=q^{-\xi}+cq^{\xi-2j}$
with $c=-1$ in (\ref{eq:g-exp-f}), is given in terms of the
positions $x_s=\onehalf[2s]_q$, $s|_{-j}^j$, of the finite
$q$-oscillator by \bea
\lambda(j-s)&=&q^{-j+s}{-}q^{-j-s}=-2e^{j\kappa}\sinh\kappa s\nonumber\\
        &=& 2q^{-j-\frac12}(q-1)\,x_s
        =-(4e^{j\kappa} \sinh\onehalf\kappa)\, x_s,
                            \label{eq:argument-Xs}
\eea
and $q=e^{-\kappa}$ as before. From (\ref{eq:g-exp-f}) thus,
the finite $q$-oscillator wave functions of mode number $n=j+m$,
$n|_0^{2j}$, are
\newpage

\bea
    \Phi^{{(2j|q)}}_n(x_s)
        &=&q^{\frac12(j+s)+\frac14 n(n-1)}
         \sqrt{\bigg[{2j\atop j+s}\bigg]_{q^2}
               \bigg[{2j\atop n}\bigg]_q
        \frac{1{+}q^{-2s}}{2(-q;q)_{2j}}}  \nonumber\\[3pt]
         & &{}\qquad{}\times K_{n}(2q^{-j-\frac12}(q-1)\,x_s;
                            \,-1,\,2j\,\vert\,q) .
                \label{eq:semi-explicit-Phi}
\eea
The explicit expression for the dual $q$-Kravchuk polynomials
in this case is
\bea
        K_{j+m}(\lambda\of{j{-}s}; -1 ,2j\,\vert\,q)
    &=&{}_3 \phi _2\left( {q^{-j-m},q^{s-j},-q^{-j-s}\atop q^{-2j},\ 0}
    \bigg |\, q;\,     q\right) \label{eq:explicit-K-poljms}\\
    &=& \sum _{k=0}^{2j}
        \frac{ (q^{-j-m};q)_k (q^{-j+s};q)_k (-q^{-j-s};q)_k}{
                (q^{-2j};q)_k} \frac{q^k}{(q;q)_k},
                        \nonumber
\eea
where $(z;q)_k$ is defined in (\ref{eq:q-bin-coeff2}).

    The lowest mode of the oscillator is [see (\ref{eq:semi-explicit-Phi})
for $n=j+m=0$],
\be
    \Phi^{{(2j|q)}}_0(x_s)=q^{\frac12(j+s)}
        \sqrt{\bigg[{2j\atop j+s}\bigg]_{q^2}
                \frac{1+q^{-2s}}{2(-q;q)_{2j}}}=\gamma^j_s.
                            \label{eq:ground-state}
\ee

    The finite $q$-oscillator wave functions possess definite parity,
\be
     \Phi^{{(2j|q)}}_n(-x_s)
        =\Phi^{{(2j|q)}}_n(x_{-s})
            =(-1)^n\, \Phi^{{(2j|q)}}_n(x_s),
                        \label{eq:parity}
\ee
and, as is to be expected, in the limit $q\to1$ return the
Kravchuk functions of the finite oscillator \cite{AW97}
\be
    \lim_{q\to 1} \Phi^{{(2j|q)}}_n(x_s)
        =  2^{-j}\sqrt{\bigg({2j\atop j+s}\bigg)
            \bigg({2j\atop n}\bigg)}\,K_n(j-s;\onehalf,2j),
                            \label{eq:limit-wavefunctions}
 \ee
 with the classical Kravchuk polynomials, introduced by Kravchuk
 in \cite{Krav}.

    The dual $q$-Kravchuk polynomials -- as all poly\-nomials -- are
analytic functions on the complex plane of their argument.  As before
in the finite oscillator models \cite{AW97,AVW99,APVW-I}, this argument
is the position coordinate, which can be analytically continued to real
or complex values $X$, even if the inner product of the space ${\cal H}_j$
is only over the point set $\{x_s\}$, $s|_{j}^j$. As to the $q$-Kravchuk
wave functions (\ref{eq:semi-explicit-Phi}) the factor in front of the
polynomial is a function that is analytic in the argument $s$ within the
interval $-j-1<s<j+1$; this means that in the position coordinate, analytic
continuation is possible within the interval $x_{-j-1}<X<x_{j+1}$.
\bigskip

\noindent
{\bf 5. Fourier-$q$-Kravchuk transform to momentum space}
\bigskip

     The identification of the position and momentum operators,
$Q=\widetilde J_1$, $P= -\widetilde J_2$ in
(\ref{eq:Q-def-twiddle})--(\ref{eq:P-def-twiddle}), brings formulae
(\ref{eq:tJpm-comm-rels}) to the role of the two Hamilton equations
(\ref{eq:Newton-Hamilton}). [This also holds for the `first' choice
using the standard basis, $Q\leftrightarrow J_1$, $P\leftrightarrow -J_2$,
that we outlined in Section 2, as well as for all oscillator models,
finite or standard.] The evolution of the finite $q$-oscillator over
time in quantum mechanics, or along the optical axis in the waveguide
model, is thus the {\it harmonic\/} motion
\be
    e^{-{\rm i}\tau H}\vecdos{Q}{P}e^{{\rm i}\tau H}
        =:\vecdos{Q\of\tau}{P\of\tau}
    =\matdos{\cos\tau}{\sin\tau}{-\sin\tau}{\cos\tau}
        \vecdos QP.             \label{eq:Hrot-QPspace}
 \ee
This is a $U(1)$ group of inner automorphisms of the \suq\
algebra, and of rotations of the phase-space surface around its
vertical axis.  It covers twice the SO(2) cycle of fractional {\it
Fourier-q-Kravchuk\/} transforms, ${\cal K}^a_q$, of power
$a=2\tau/\pi$ and angle $\tau$, \be
     {\cal K}^a_q :=
        \exp({-{\rm i}\,\pi a\,(J_3+j)/2})
        =e^{{\rm i}\,\pi a/4}\,\exp({-{\rm i}\,\pi a H /2}).
                        \label{eq:FK-phase-oscil}
\ee

    For $a=1$ we have the Fourier-$q$-Kravchuk transform
${\cal K}_q$. The action of ${\cal K}_q$ on the eigenbasis of
position yields the eigenbasis of momentum,
\be
    \widetilde g^j_s\of{x}:= {\cal K}_q\,g^j_s\of{x}.
                        \label{eq:K-acton-g}
\ee
These functions have the properties and form
\bea
    P\,\widetilde g^j_r\of{x}
        &=& -Y_r\,\widetilde g^j_r\of{x}, \quad
                Y_r=\onehalf [2r]_q
                =\frac{\sinh r\kappa}{2\sinh\frac12\kappa}
                =-Y_{-r},\quad r|_{-j}^j,
                        \label{eq:eigenv-eq-J2}\\
        \widetilde g^j_r\of{x} &=& g^j_r\of{{\rm i}x}
                        =g^j_{-r}\of{{-{\rm i}x}} \\ \nonumber
                &=& \gamma^j_r\,
            ({\rm i}q^{\frac14(1-2j)} x;q)_{j-r}\,
            (-{\rm i}q^{\frac14(1-2j)} x;q)_{j+r},
                        \label{eq:e-eq-tildeg}
 \eea
 where $\gamma^j_r$ is the constant given in
(\ref{eq:c-in-eq}); the spectrum of momenta, $Y_r$, $r|_{-j}^j$,
is the same as that of position [{\it cf}.
(\ref{eq:eigenv-eq-J1})]. Since ${\cal K}_q^a$ is unitary under
the inner product in ${\cal H}_j$, the Fourier-$q$-Kravchuk
transform of the finite $q$-oscillator eigenfunctions
(\ref{eq:def-of-Phi})--(\ref{eq:semi-explicit-Phi}) of modes
$n=j+m$, are \be
    \widetilde\Phi^{{(2j|q)}}_n(x_s)
    := {\cal K}_q\, \Phi^{{(2j|q)}}_n(x_s)
            :=(g^j_s,{\cal K}_q\,f^j_m)_{{\cal H}_j}
                =(-{\rm i})^{n}\Phi^{{(2j|q)}}_n(x_s),
                        \label{eq:K-acton-PhiXs}
\ee
as in all oscillator models.

    The Fourier-$q$-Kravchuk transform of a function or signal
$\Phi$, of values $\Phi(x_s)=(g^j_s,\Phi)_{{\cal H}_j}$ on the
finite, discrete sensor point set $\{x_s\}$, $s|_{j}^j$, is
defined by
\be
    \widetilde\Phi(x_r)=(\widetilde g^j_r,\Phi)_{{\cal H}_j}
            =\sum_{s=-j}^j K^{(2j|q)}_{r,s}\,\Phi(x_s),
                            \label{eq:FqK-ofarb-funct}
\ee where the kernel is the overlap of the position eigenfunctions
$g^j_s$ in (\ref{eq:eigenf-eq-g}) with the momentum eigenfunctions
$\widetilde g^j_r$ in (49),
 \be
    K^{(2j|q)}_{r,s}:=(\widetilde g^j_r,g^j_s)_{{\cal H}_j} .
                        \label{eq:direct-df-qFTkernel}
\ee
This kernel is given explicitly below in (\ref{eq:FqK-tfmn-kernel})
with $a=1$.

\bigskip

\noindent
{\bf 6. Fractional Fourier-$q$-Kravchuk kernel}
\bigskip

    The Fourier-$q$-Kravchuk transform (\ref{eq:K-acton-PhiXs})
is fractionalized by the operator ${\cal K}_q^a$ in
(\ref{eq:FK-phase-oscil}), independently of the realization,
on the mode eigenbasis of $J_3$,
\be
    {\cal K}_q^a\,f^j_m=\exp({-{\rm i}\,\pi a(j+m)/2})\,f^j_m
                        =\exp({-{\rm i}\,\pi n a/2})\,f^j_m.
                        \label{eq:K-acton-fs}
\ee
When we apply ${\cal K}_q^a$ on a finite, complex `signal'
function of $2j+1$ points,
\be
    \Phi(x_s)=(g^j_s,\Phi)_{{\cal H}_j}
            =\sum_{m=-j}^j (g^j_s,f^j_m)_{{\cal H}_j}
                    (f^j_m,\Phi)_{{\cal H}_j}
                            \label{eq:arb-funct}
\ee
we obtain another such function, labelled by $a$,
$$
    \Phi^\oof{a}(x_s):={\cal K}_q^a\, \Phi(x_s)
                :=(g^j_s,{\cal K}_q^a\, \Phi)_{{\cal H}_j}
                =({\cal K}_q^{-a}\, g^j_s,\Phi)_{{\cal H}_j}
$$   \be
        =\sum_{m=-j}^j ({\cal K}_q^{-a}\, g^j_s,f^j_m)_{{\cal H}_j}
                    (f^j_m,\Phi)_{{\cal H}_j}
        =\sum_{m=-j}^j (g^j_s,{\cal K}_q^{a}\, f^j_m)_{{\cal H}_j}
                    (f^j_m,\Phi)_{{\cal H}_j}  \label{eq:FqK-tfmof-Phi}
\ee
$$
        =\sum_{m=-j}^j e^{-{\rm i}\,\pi a(j+m)/2}
                (g^j_s,f^j_m)_{{\cal H}_j}
            \sum_{s'=-j}^j (f^j_m,g^j_{s'})_{{\cal H}_j}
                        (g^j_{s'},\Phi)_{{\cal H}_j}
        = \sum_{s'=-j}^j K_{s,s'}^{(a,2j|q)}
                        \Phi(x_{s'}),
$$
where the fractional Fourier-$q$-Kravchuk transform kernel
$K_{s,s'}^{(a,2j|q)}$ is a $(2j{+}1)\times(2j{+1})$ matrix of
elements given by the bilinear generating function
\cite[formula (8.15)]{JeugtJag}
 \be
    {}\hskip-1cm{} K_{s,s'}^{(a,2j|q)}
        :=\sum_{n=0}^{2j}
                \Phi^{{(2j|q)}}_n(x_s)\,
                        e^{-{\rm i}\,\pi na/2}\,
                \Phi^{{(2j|q)}}_n(x_{s'})^*
                            \label{eq:FqK-tfmn-kernel}
\ee  \be
       = \gamma^j_s\,\gamma^j_{s'}\,\beta_{s,s'}(t){}
      {}_8{}W_7(-q^{-2j-1}t;\; q^{s-j},-q^{-j-s},
                q^{s'-j},-q^{-j-s'},-t;\ q,-t), \label{eq:FqK-W-phi-funct}
\ee where
\be
 t:=e^{-{\rm i}\, \pi a/2}, \label{eq:t-and-cs}
\ee
\be
         \beta_{s,s'}(t):=
    \frac{(q^{s-j}t,-q^{-j-s}t,q^{s'-j}t,-q^{-j-s'}t,-t;q)_\infty
    }{(q^{s-s'}t,-q^{s+s'}t,q^{s'-s}t,-q^{-s-s'}t,-q^{-2j}t;q)_\infty},
                            \label{eq:def-asst}
\ee
and $\gamma^j_s$ is given by (\ref{eq:c-in-eq}). The function
${}_8{}W_7$, defined in  \cite{GasRah}, is
\be
    {}_8{}W_7(a;b,c,d,e,f;\ q,z)
    :=\sum_{k=0}^\infty
        \frac{1-aq^{2k}}{1-a}
        \frac{(a,b,c,d,e,f;q)_k\, z^k}{(q,qa/b,qa/c,qa/d,qa/e,qa/f;q)_k},
                                \label{eq:basic-hypfunction}
\ee
where $(a,\ldots,c;q)_\infty:=(a;q)_\infty\*\ldots\*(c;q)_\infty$
and $(a;q)_\infty=\prod_{k=0}^\infty(1-aq^k)$ in accordance with
(\ref{eq:q-bin-coeff2}). This function can be expressed in terms
of the basic hypergeometric function ${}_8\phi_7$ (see
\cite[\S2.2, formula (2.5.1)]{GasRah}), with coefficients which
allow it to be reduced to the basic hypergeometric function
${}_4\phi_3$:
$$
    {}_8{}W_7(-q^{-2j-1}t;\; q^{s-j},-q^{-j-s},
                q^{s'-j},-q^{-j-s'},-t;\ q,-t)
$$ \be
    = \frac{(-q^{-2j}t,q^{-j-s'},-q^{-j+s'},t;q)_{\infty}}
              {(-q^{-j-s'}t,q^{-j+s'}t,q^{-2j},-t;q)_{\infty}}
        {}_4{}\phi_3\left( {q^{-j+s'}, -q^{-j-s'},t,-t\atop
            -q^{-j-s}t,q^{-j+s},-q}\; \bigg |\,  q,q\right).
                        \label{eq:basic-4F3}
\ee
 We also note that due to relation $(a;q)_n=(a;q)_\infty
/(aq^n;q)_\infty$, the expression for $\beta_{s,s'}(t)$ in
(\ref{eq:def-asst}) can be reduced to \be
    \beta_{s,s'}(t)=\frac{(q^{s-j}t;q)_{j-s'} (q^{-j+s'}t;q)_{j-s}
        (-q^{-j-s'}t;q)_{j-s} (-q^{-j-s}t;q)_{j+2s+s'}}{
            (-q^{-2j}t;q)_{2j}}.
                        \label{eq:alternative-beta}
\ee

     Naturally, ${\cal K}_q^{a_1}\,{\cal K}_q^{a_2} ={\cal
K}_q^{a_1+a_2}$ and ${\cal K}_q^0= \hat1$. The `phase correction'
by $\pi a=2\tau$ which we introduced in (\ref{eq:FK-phase-oscil})
implies that ${\cal K}_q^4=\hat1$ (as the ordinary Fourier
integral transform), while the fourth power of the oscillator
evolution operator $\exp({\rm i}\tau H)$ is $-\hat1$ for the full
rotation angle $\tau=2\pi$. This is the analogue of the
metaplectic sign of the waveguide case (see \cite{SimW}, {\it
cf}.\ \cite{Ozak}), where the U(1) subgroup generated by the
latter covers twice the SO(2) of the former. Parity is conserved
under the fractional Fourier-Kravchuk transformation because $J_3$
commutes with the inversion of phase space. And again, in the
limit $q\to1$ we recover the previous Fourier-Kravchuk kernel
expressed in terms of the Wigner `little-$d$' functions
\cite{AVW99}, \bea
    \lim_{q\to1}K_{s,s'}^{(a,2j|q)}
    =  K^{(a,2j)}_{s,s'}
         = e^{-{\rm i}\,\pi j a/2 }\,(-{\rm i})^{s-s'}\,
                        d^{j}_{s,s'}(\onehalf\pi a).
                    \label{eq:Frac-Fou-Kravchuk-kernel}
\eea

\bigskip

\noindent
{\bf 7. Equivalent potentials}
\bigskip

    In ordinary quantum mechanics, the ground state $\Psi_0\of{x}$
of a system with a potential $V\of{x}$ and energy $E_0>-\infty$, has
no zeros; thus, the Schr\"odinger equation determines the potential
energy of the system from the ground state,
\be
    \left( -\onehalf\totder{^2}{x^2}+V\of{x}-E_0\right) \Psi_0\of{x}=0
        \quad\Rightarrow\quad V\of{x}-E_0
            = \onehalf\totder{^2}{x^2}\Psi_0\of{x}\Big/\Psi_0\of{x}.
                    \label{eq:Schroedeq-implies-potential}
\ee
As a well-known example we have the harmonic oscillator, whose
ground state is $\Psi_0\of{x}\sim e^{-\frac12 x^2}$, so
$\totder{^2}{x^2}\Psi_0\of{x}=(x^2-1)\Psi_0\of{x}$ and
(\ref{eq:Schroedeq-implies-potential}) yields correctly
$V\of{x}-E_0=\onehalf x^2-\onehalf$.

    In the case when the system is discrete over the set of points
$x_s=sh+x_0$, with integer $s$, which are equidistant by $h$, an
equivalent potential may be defined following
(\ref{eq:Schroedeq-implies-potential}). We qualify it as {\it
equivalent\/} because the discrete systems, that have been studied
(such as Kravchuk, Meixner and Hahn systems
\cite{AtSus}--\cite{AtJaf}, \cite{APVW-II,APWcontr-rad}), obey
Schr\"odinger-type difference equations which do {\it not\/}
separate into a sum of terms, where one is readily identifiable
with the kinetic term of the second-degree difference operator,
plus a potential term that is only dependent on position $x_s$.
The symmetric second-difference operator, acting on functions of
$x_s$, can be expressed in terms of the right-difference and the
left-difference operators $\nabla_\ssr{\!R}$ and
$\nabla_\ssr{\!L}$, \be
      \begin{array}{rcl}
    \nabla_\ssr{\!R}&:=&\displaystyle\frac{\Delta}{\Delta x_s}
            =\frac1{\Delta x_s}(e^{\partial_s}-1)
            =\frac1{x_{s+1}-x_s}(e^{\partial_s}-1),\\
    \nabla_\ssr{\!L}&:=&\displaystyle\frac{\nabla}{\nabla x_s}
            =\frac1{\nabla x_s}(1-e^{-\partial_s})
            =\frac1{x_s-x_{s-1}}(1-e^{-\partial_s}),\end{array}
                    \label{eq:nabladelta}
\ee
where $\Delta=e^{\partial_s}-1=e^{\partial_s}\nabla$. So, a
difference analogue of the differential operator $d^2/dx^2$
in (\ref{eq:Schroedeq-implies-potential}) has the form
\be
    \frac1{x_{s+1/2}-x_{s-1/2}}(\nabla_\ssr{\!R}-\nabla_\ssr{\!L}).
                    \label{eq:seventy-three}
\ee
Consequently, when the ground state of the system is
$\psi\of{s}:=\Psi_0\of{x_s}$, the equivalent potential,
according to its quantum-mechanical correspondent in
(\ref{eq:Schroedeq-implies-potential}), is
\bea
     & &   V\of{x_s}-E_0
         = \displaystyle \frac1{2\psi\of{s}[x_{s+1/2}-x_{s-1/2}]}
                (\nabla_\ssr{\!R}-\nabla_\ssr{\!L})\psi\of{s}
                \label {eq:left-right-derivatives}\\
      & &{}\qquad{}=\displaystyle\frac1{2(x_{s+1/2}-x_{s-1/2})\,\psi\of{s}}
            \left( \frac{\psi\of{s{+}1}-\psi\of{s}}{x_{s{+}1}-x_{s}}
                  -\frac{\psi\of{s}-\psi\of{s-1}}{x_{s}-x_{s-1}} \right).
                    \nonumber
\eea

    In the case of the finite Kravchuk oscillator, the set of values
of position $x_s=s$ ($h=1$ and $x_0=0$) is finite: $\{x_s\}_{s=-j}^j$.
Yet, the wave functions $\psi\of{s}:=\Psi_0^{(2j)}\of{x_s}$ can be
analytically continued in $x$ everywhere except for branch-point zeros
at $x_{\pm(j{+}1)}:=\pm(j{+}1)$, which are due to the square root of
the binomial distribution. Thus, on the closed segment
$x_{-(j{+}1)}\le x\le x_{j{+}1}$, the second difference in
(\ref{eq:left-right-derivatives}) is defined for any real value of
$x$ in the interval $x_{-j}\le x\le x_{j}$. A similar extension
and range of validity holds for the Meixner and Hahn oscillator
cases \cite{AtJaf,APVW-II,APWcontr-rad}. The lowest mode of the
Kravchuk oscillator, where $h=1$, is given in (\ref{eq:ground-state}).
>From this one derives the equivalent potential for the Kravchuk
eigenfunction system
\bea
       V\of{x_s}-E_0+1
    &=& \displaystyle\frac{\psi\of{s+1}+\psi\of{s-1}}{2\psi\of{s}}
    \nonumber\\
        &=& \displaystyle\frac{\sqrt{(j+s)(j+s+1)}+\sqrt{(j-s)(j-s+1)}}{
                2\sqrt{(j+1)^2-s^2}}.
                \label{eq:Kravchuk-potential}
\eea

   When the set of position values is {\it not\/} equally spaced, as is
the case in the finite $q$-oscillator, $\{x_s\}_{s=-j}^j$ as in
(\ref{eq:eigenv-eq-J1}), we shall consider the differences with
respect to the position coordinate
\be
    x_s=\onehalf[2s]_q
    =\frac{\sinh s\kappa}{2\sinh\onehalf\kappa}\quad\Rightarrow\quad
    \left\{  {x_{s+1}-x_s=\cosh(s{+}\onehalf)\kappa,\atop
     x_{s}-x_{s-1}=\cosh(s{-}\onehalf)\kappa.}\right.
            \label{eq:Xss-dfferences}
\ee
Taking into account that
\be
    \psi(s+1)= q^{-s-1/2}
        \sqrt{ \frac{\cosh(s+1)\kappa}{\cosh s\kappa}
               \frac{\sinh(j-s)\kappa}{\sinh (j+s+1)\kappa}}\,
                \psi\of{s}, \label{eq:phis+1}
\ee
we arrive at the expression for the equivalent potential in
the general case
\bea
    V\of{x_s}-E_0&=&
    \frac1{2\,q^{1/2}\,\cosh(s+\onehalf)\kappa\,\cosh(s-\onehalf)\kappa}
         \nonumber\\
         & &\times\bigg\{
    q^{s}\, \frac{\cosh(s+\onehalf)\kappa}{\cosh s\kappa}
        \sqrt{ \frac{\cosh(s-1)\kappa}{\cosh s\kappa}
               \frac{\sinh(j+s)\kappa}{\sinh (j-s+1)\kappa}}
               \nonumber\\
             & & +
    \, q^{-s}\, \frac{\cosh(s-\onehalf)\kappa}{\cosh s\kappa}
        \sqrt{ \frac{\cosh(s+1)\kappa}{\cosh s\kappa}
               \frac{\sinh(j-s)\kappa}{\sinh (j+s+1)\kappa}}
               \nonumber\\
             & &\qquad{}
                - 2\,q^{1/2}\,\cosh\onehalf\kappa\bigg\}
                            \label{eq:first-V0}
\eea
for functions $\psi(s):=\Psi_0^{(2j|q)}(x_s)$ (see formula (42)).
Obviously, in the limit when $q\to1$ (that is, $\kappa\to0$), this
expression coincides with (\ref{eq:Kravchuk-potential}).

Note that acceptable ground states occur for values of $q$ which
are larger of some number $a<1$ (this number $a$ changes with the
value of $j$) while lower values of $q$ present the raised-wings
problem of interpretation. The corresponding potentials have an
oscillator-type form for all values of $q$ and this property is of
course likewise shared by the $q$-Kravchuk wave functions. A study
of these functions with attention to their oscillations and
convergence should be undertaken but this task is beyond the
purpose of the present paper.

\bigskip

\noindent {\bf 8. Contraction of the algebra $\suq\rightarrow{\it
osc}_q$}
\bigskip

     We consider a sequence of finite $q$-oscillators over sets of $2j+1$
points which increase in number and density as $j\to\infty$, while the
mode number $n=j+m$ remains finite, \ie, near to the ground state $n=0$
(for eigenvalues $m$ of $J_3$ near to $-j$). The spectrum of the Hamiltonian
operator $H=J_3+j+\onehalf$ of the $q$-oscillator retains its linear
lower-bound spectrum (\ref{eq:position-energy-spectra2}) for all $j$'s in
the sequence. In the case of the ($q=1$) finite oscillator, we showed in
\cite{APWcontr} that the ${\rm su}(2)$ dynamical algebra, wave
functions, and Fourier-Kravchuk transform, contract to the ordinary
oscillator algebra ${\it osc}=\hbox{span}\,\{Q,P,H,\hat1\}$. In the
present $q$-case we follow an analogous contraction to the $q$-oscillator
model of Macfarlane and Biedenharn \cite{Macfar,Bieden}; nevertheless,
there are some important differences between the $q$- and non-$q$ cases
that we shall point out below.

    The `sensor points' of our finite $q$-oscillator [\ie, the spectrum
of $Q\in{\rm su}_q(2)$, $\Sigma(Q)$ in (\ref{eq:position-energy-spectra1})]
extend between $x_{-j}$ and $x_j$, inside an interval which grows
asymptotically with $j$ as $\sim q^{-j}= e^{j\kappa}$ (for $0<q=e^{-\kappa}<1$,
$\kappa>0$) --- and are not equally-spaced within. Our contraction process
will keep the range of positions finite by introducing, for each finite $j$,
the operators
\be
    Q^{(j)}:= w_j\,Q,\qquad P^{(j)}:= w_j\,P,
                \label{eq:def-Qj-Pj}
\ee
scaled with coefficients whose asymptotic behavior is appropriate,
\be
    w_j:=\frac{q^{\frac12(j+\frac12)}}{\sqrt{x_j}}
        =e^{-\frac12(j+\frac12)\kappa}
            \sqrt{\frac{2\sinh\frac12\kappa}{\sinh j\kappa}}
        \sim q^j\sqrt{2(1-q)}
            =e^{-j\kappa}\sqrt{e^{-\frac12\kappa}\sinh\onehalf\kappa}.
                                \label{eq:def-wj}
\ee
The {\it number\/} operator, $N:=H-\onehalf=J_3+j$, is assumed to act
on a subspace of functions whose mode eigenvalues $n=j+m$ remain
finite in $n\in\{0,1,\ldots\}$.

    As we let $j\to\infty$, the \suq\ algebra of the finite
$q$-oscillator will contract to a different $q$-algebra, that will
characterize the `continuous' limit of our finite model.
The commutation relations (\ref{eq:tJpm-comm-rels}), which can be
written
\be
    [H,Q^{(j)}]=-{\rm i}\,P^{(j)},\qquad
       [H,P^{(j)}]= {\rm i}\,Q^{(j)},
                \label{eq:contract-tJpm-comm-rels}
\ee
continue to be harmonic oscillator Hamilton equations. The third
commutator (\ref{eq:def-J3twiddle}), which is characteristic of our
\suq\ finite model, becomes
\be
    [Q^{(j)},P^{(j)}]=w_j^2[Q,P]
        = {\rm i}\,\frac{q^{(j+1/2}}{x_j} F_q(C_q,J_3).
                \label{eq:contract-third}
\ee
Acting on the subspace of functions whose mode numbers remain finite,
from (\ref{eq:widetilde-J-diagonal}) we find that the asymptotic
behavior of the right-hand side of (\ref{eq:contract-third}) is
\be
    \frac{q^{(j+\frac12)}}{x_j} F_q(C_q,J_3)
        \sim q^{J_3+j}=q^{H-\frac12}=q^N.
                    \label{eq:asympt-J3-H}
\ee
When $j\to\infty$, the formal limit operators $Q^{(j)}\to\overline{Q}$
and $P^{(j)}\to\overline{P}$ satisfy the oscillator Hamilton equations
(\ref{eq:contract-tJpm-comm-rels}) and
\be
    [\overline{Q},\overline{P}]={\rm i}\,q^N,\qquad N=H-\onehalf.
                    \label{eq:new-comm-rel}
\ee

    The reader may be more familiar with the contracted algebra
$\hbox{span}\,\{\overline{Q}$, $\overline{P},N\}$ when it is
written in terms of the raising and lowering operators as \be
    A_\pm:=\overline{Q}\mp {\rm i}\,\overline{P}
        =\lim_{j\to\infty}\widetilde J_\pm,
                    \label{eq:Aplusminus}
\ee
whose commutation relations are
\be
    [A_+,A_-]=2q^N,\qquad A_-\,A_+ - q\, A_+\,A_-=\hat1.
                            \label{eq:qq-comm-rels}
\ee
This we identify as the $q$-oscillator algebra ${\it osc}_q$
defined by Macfarlane \cite{Macfar} and Biedenharn \cite{Bieden}.
The $j\to\infty$ limit of (\ref{eq:Aplusminus}) yields
\bea
    A_+\,\Psi^{(q)}_n(X) &=&  \sqrt{\{ n+1\}_q}\Psi^{(q)}_{n+1}(X),
                        \label{eq:ordinary-qosc-1}\\
    A_-\,\Psi^{(q)}_n(X) &=& \sqrt{\{ n\}_q }\,\Psi^{(q)}_{n-1}(X),
                        \label{eq:ordinary-qosc-2}
\eea
where $\{n\}_q:=(q^n-1)/(q-1)$  and
\be
    \Psi^{(q)}_n(X) =\frac1{\sqrt{n!}}\,(A_+)^n\,\Psi^{(q)}_0(X)
                        \label{eq:ordinary-qosc-3}
\ee
are mode eigenfunctions obtained from $A_-\Psi^{(q)}_0(X)=0$.

    We would like to point out however, that before the limit
$j\to\infty$ is achieved, the spectra of position and momenta,
(\ref{eq:eigenv-eq-J1}) and (\ref{eq:eigenv-eq-J2}), are
asymptotically constrained to a {\it finite\/} position interval
\be
    |\Sigma(Q^{(j)})|\le w_jx_j \sim 1/\sqrt{2(q^{-1}-1)}.
                        \label{eq:finite-contr-lim}
\ee
Only in the $q=1$ finite oscillator case \cite{APWcontr}, where $x_j=j$,
does the position interval grow to the real line as $\sim \sqrt{j}$,
keeping equal distances $\sim1/\sqrt{j}$ between neighboring sensor
points. For any other $0<q<1$, all points $x_s$ of $\Sigma(Q^{(j)})$
except $x_{\pm j}$, will crowd towards zero in the middle of the
interval. This feature of the contraction limit between $q$-algebras
is at variance with that encountered with Lie algebras, where one can
extend the operation from formal operators to finite Hilbert spaces
of growing dimensions, to find limits from Kravchuk to Hermite
functions, and Schr\"odinger difference to differential equations.
This matter also requires a separate, deeper analysis that we leave
for a separate publication.

\bigskip

\noindent
{\bf Acknowledgements}
\bigskip

We thank the support of the Direcci\'on General de Asuntos del
Personal Acad\'emico, Universidad Nacional Aut\'onoma de M\'exico
({\sc dgapa--unam}) by the grant IN102603-3 {\it \'Optica
Matem\'atica\/} and {\sc sep-conacyt} project 41051-F. A.U.K.\
acknowledges {\sc conacyt} (M\'exico) for a C\'atedra Patrimonial
Nivel II.

\end{document}